\documentclass{pasj01}
\usepackage[switch,mathlines]{lineno}
\Received{$\langle$03-Nov-2023$\rangle$}
\Accepted{$\langle$18-Dec-2023$\rangle$}
\Published{$\langle$publication date$\rangle$}
\usepackage{color}
\newif\ifMARKED
\MARKEDfalse  % 出版用

\ifMARKED
    \newcommand{\ch}[1]{\textcolor{red}{#1}}
\else
    \newcommand{\ch}[1]{#1}
\fi

\begin{document}

\title{Survey of public attitudes toward astronomy in Japan}
\author{Naohiro Takanashi\altaffilmark{1,}\footnotemark[*], Masaaki Hiramatsu\altaffilmark{2,3}, Shio Kawagoe\altaffilmark{4}, Nobuhiko Kusakabe\altaffilmark{2,5}, Koki Sawada\altaffilmark{6}, and Harufumi Tamazawa\altaffilmark{4}}%
\altaffiltext{1}{Executive Management Program, the University of Tokyo, Bunkyo-ku, Tokyo 113-8654, Japan}
\altaffiltext{2}{National Astronomical Observatory of Japan, Mitaka, Tokyo 181-8588, Japan}
\altaffiltext{3}{The Graduate University for Advanced Studies, SOKENDAI,  Mitaka, Tokyo, 181-8588, Japan}
\altaffiltext{4}{Institute of Industrial Science, the University of Tokyo, Meguro-ku, Tokyo 153-8505, Japan}
\altaffiltext{5}{Astrobiology Center, National Institutes of Natural Sciences, Mitaka, Tokyo 181-8588, Japan}
\altaffiltext{6}{Graduate School of Tourism, the University of Wakayama, Wakayama, Wakayama 640-8510, Japan}
\email{naohiro.takanashi@emp.u-tokyo.ac.jp}


\maketitle

\begin{abstract}
We report on the results of a survey we conducted on the Japanese public's attitudes toward astronomy. This survey was conducted via an online questionnaire, with 2,000 responses received. Based on this data, we present what kind of interest the general public in Japan has in astronomy. We also conducted a questionnaire survey of those involved in astronomy communication to examine how they differ from the general public. The results suggest that while there are clear differences \ch{between them} in terms of their engagement in astronomy, there is also continuity between them by looking at their attributes in more detail. \ch{The data presented in this paper could help us to promote communicating astronomy to the public.}

\end{abstract}
%\pagewiselinenumbers

\section{Introduction}

As discussed in the International Astronomical Union\footnote{i.e., IAU Strategic Plan 2020-2030}, discussing good relationships between astronomy and the public is extremely important for the sustainable development of astronomy in the future. Not only researchers but also educators, professional communicators and citizen communicators are also engaged in various activities to connect astronomy and the public. You can easily find many interesting reports of the activities on journals such as Communicating Astronomy with the Public (CAP) Journal and so on. 

Quantitative studies have been conducted to examine attitudes of the people involved in those activities. The Astronomical Society of the Pacific discussed the potential for informal education of astronomy by using the power of amateur astronomers in US, after revealing their realities through a web-based survey (cf. \cite{Bennett2004}). Based on the responses of 744 respondents attending astronomy and space outreach events in UK, \citet{Entradas2014} discussed how we can reach new audiences who are not often targeted by conventional outreach efforts. \citet{Dang2015} reported astronomers' views on public outreach examined through an online and in-person survey of 155 respondents. \citet{Marusic2018} showed students' attitudes towards astronomy based on a questionnaire survey with 396 students of both Bosnia and Herzegovina and Croatia. \citet{ishii2020} presents the results of a survey of how students (n=170) in education departments specializing in the sciences perceive space science and astronomy. It can be said that these studies quantify the outcomes of the activities that connect astronomy and society from various perspectives through the surveys.

On the other hand, there are few studies which show the attitudes of the general public, who are not necessarily involved in such activities, toward astronomy. We can find studies which show the attitudes of the general public toward space exploration (i.e., \cite{Smith2003,Pfleger2022}), science in general (i.e., \cite{Ataka2008,BEIS2019}), however, we can not find those toward astronomy. Despite the importance of discussing the relationship between astronomy and society based on objective data, efforts to obtain such data have been insufficient.

In Japan, one of the pioneering efforts to show the attitudes of the general public is the "Handbook of Public Observatories," which was compiled by Nishi--Harima Astronomical Observatory\footnote{Nishi--Harima Astronomical Observatory $\langle http://www.nhao.jp/\rangle$} in 1992, 1993 and 1997. Originally, this handbook was created to promote horizontal cooperation and information exchange among public observatories. Later, as a successor to this handbook, the "White Paper on Public Observatories" compiled by Japan Public Observatory Society (JAPOS)\footnote{JAPOS $\langle https://www.koukaitenmondai.jp/\rangle$} was published in 2006, which includes not only data but also analysis of the data and proposals based on the analysis. Similar data collections is the "Planetarium White Paper" compiled by the Japan Planetarium Society (JPS) (2001, 2005) and its successor, the "Planetarium Data Book" compiled by the Japan Planetarium Association (JPA)\footnote{JPA $\langle https://planetarium.jp/\rangle$}. It should be noted, however, that while these data were intended to explore trends among the general public, the survey was focused on understanding the facility and its operations.

There are also some surveys targeting the Japanese general public. \citet{hayakawa2013} reported the effects of annular solar eclipse on public awareness of science and technology through a web-based survey (n = 1,600). A survey of citizens' attitude to astro--tourism conducted by the Sora Tourism Promotion Council (Agata 2019, unpublished), a survey toward those who are willing to financially support astronomy (or space science) conducted by the Japan Aerospace Exploration Agency (JAXA)\footnote{JAXA $\langle https://www.jaxa.jp/\rangle$} (Ikuta 2020, unpublished), and a survey of awareness of the Atacama Large Millimeter/submillimeter Array (ALMA) conducted by the National Astronomical Observatory of Japan (Hiramatsu 2021, unpublished) are examples of such surveys in Japan. While we can get some insights of the public's attitude toward astronomy from these surveys, it is not easy to get the whole picture since these surveys are designed to put a focus on specific purposes.

In this study, we present attitude toward astronomy in Japan as a result of online survey of general public. We cannot find any similar surveys in the field of astronomy in Japan. After understanding the general trend, we also conducted a similar survey toward those who involve in astronomy education and public outreach in Japan to see how their attitudes toward astronomy differ from those of the general public.

\section{Surveys}
%一般市民向け、関係者向け
Surveys were conducted both with the general public and with those involved in astronomy education and public outreach. Hereafter, we call the latter people "EPO people" in this paper. Both surveys were conducted in Japanese.

For the general public, an online survey was conducted through Cross Marketing, Inc., which is a research company in Japan, over a four--day period from February 17 to 20, 2023, and responses were received from 2,000 people (male = 992, female = 996, others = 12). \ch{Respondents were people aged 20-69, selected from monitors registered with the survey firm, and were assigned according to the demographic and regional distribution of Japan.} They were asked about the full set of items described below.

For the EPO people, an online survey was conducted by ourselves over from February 22 to 28, 2023, and responses were received from 178 people (male = 113, female = 64, others = 1). We invited people to respond to the survey through the mailing lists of Japanese Society for Education and Popularization of Astronomy (JSEPA)\footnote{JSEPA $\langle https://www.tenkyo.net\rangle$}, JAPOS, JPA and the Star Sommeliers\footnote{People active in star gazing events for the public in Japan}. Respondents consist of researchers (astronomy, education, science communication, etc.), professional communicators (teachers, staff of planetariums, science centers and public observatories, etc.), citizen communicators (amateur astronomers, volunteers, etc.), and others (active participant). They were asked about only some of the items in order to increase the response rate. \ch{We do not know the exact response rate due to overlap in membership of each organization, but we estimate that the response rate is about 10\%.}

\subsection{Procedure}

\subsubsection{Survey of General Public}

The survey items for the general public consist of (1) sociodemographic variables and (2) questionnaire items.

The sociodemographic variables are AGE, SEX (male, female and others), LOCATION (prefecture), MARRIAGE (married or not), CHILDREN (have a child or not), EDUCATION (educational background) and OCCUPATION. These survey items were not set by us, but were provided as basic information by the survey company.

Number of the questionnaire items are 28 in total. 13 of the items were prepared for other research purposes and are not addressed in this paper. The remaining 15 items (Q1-Q15) are listed in Appendix 1. The first three items (Q1-Q3) are prepared for measuring degree of interest for science and technology based on the Victorian Segmentation (VSEG) method \citep{VSEG}, which segment the general public and discuss their property at a higher resolution. The Japanese translation followed the method of \cite{kano2013}. The remaining 12 items (Q4-Q15) are prepared for measuring degree of interest for astronomy from several different point of views.

Many of these questionnaire items were newly proposed for this survey from a point of view toward comparison with prior surveys and researches. In particular, we were conscious of preparing items that might show differences compared to other countries.

\subsubsection{Survey of EPO People}

The survey items for the EPO people consist of (1)  sociodemographic variables and (2) questionnaire items. Comparing to the survey for the general public, we reduce the number of items. 

The sociodemographic variables are only AGE, SEX (male, female and others), LOCATION (prefecture) and POSITION (how to get involved with astronomy). 

The questionnaire items are nine in total. Four of the items were prepared for other research purposes and are not addressed in this paper. The remaining five items (Q$_{E}$1-Q$_{E}$5) are also listed in Appendix 1. The first three items (Q$_{E}$1-Q$_{E}$3) are prepared for measuring degree of interest for science and technology based on the Victorian Segmentation (VSEG) method, while the remaining two items (Q$_{E}$4 and Q$_{E}$5) asked about the experiences. In order to compare with the results by general public, the texts of the questionnaire items are identical in both surveys.

\section{Analysis}

The collected data were analyzed to examine correlations among the three clusters: sociodemographic variables, interest in science and technology (Q1-Q3, Q$_{E}$1-Q$_{E}$3), and interest in astronomy (Q4-Q15, Q$_{E}$4-Q$_{E}$5).

First, we examined the relationship between sociodemographic variables and interest in science and technology, which can be comparable to previous studies. Then, we analyzed how different trends in interest in astronomy appear based on the level of interest in science and technology. The former analysis was performed on survey data for the general public only, while the latter was performed on the survey data for both general public and the EPO people.

Social surveys generally involve observational bias. Our data may also involve biases caused by various reasons such as the survey procedure. We are aware of these biases and will discuss the impact of them on the results later.

\section{Results}

\subsection{Overview}

Here is a general overview of the data we obtained. We present the results of both the survey for the general public and the results of the survey for the EPO people. All data are available online\footnote{https://survey.tenpla.net}.

\subsubsection{Survey of General Public}

%socio
As we described, the respondents were assigned according to the demographic and regional distribution of Japan. Average age of the respondents is 45.9 years old. 57.8\% of the respondents are married (or were married) and the remaining 42.1\% are not married. 56.7\% of the respondents have no children and the remaining 43.3\% have children. The most common occupation of respondents is working for a company, with 39.9\% of respondents being full time workers. The next highest percentage was part--time workers at 15.5\%, followed by homemaker at 13.1\%. 47.5\% of respondents are graduates of university or graduate school, while 52.5\% are graduates of other schools (elementary / junior high / high school, junior college, vocational school). These figures are generally consistent with data published by the Statistics Bureau of Japan's Ministry of Internal Affairs and Communications \footnote{https://www.stat.go.jp/}.

\begin{figure}[hbtp]
 \centering
 \includegraphics[width=80mm]
      {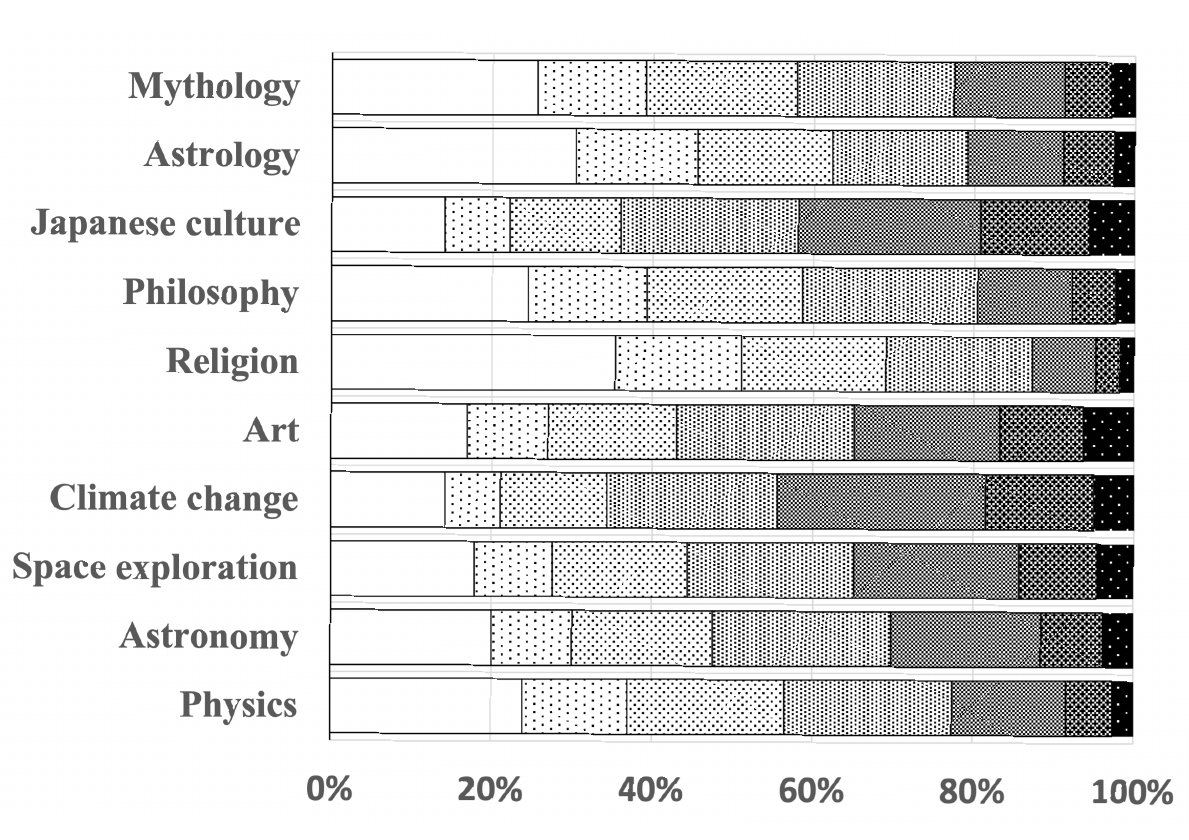}
\raggedright
\caption{Response of Q4, "How interested are you in the items?" From left to right: "not interested at all," "not interested," "not very interested," "neither interested nor disinterested," "a little interested," "interested," "very interested."}
 \label{Q4}
\end{figure}

\begin{figure}[hbtp]
 \centering
 \includegraphics[width=80mm]
      {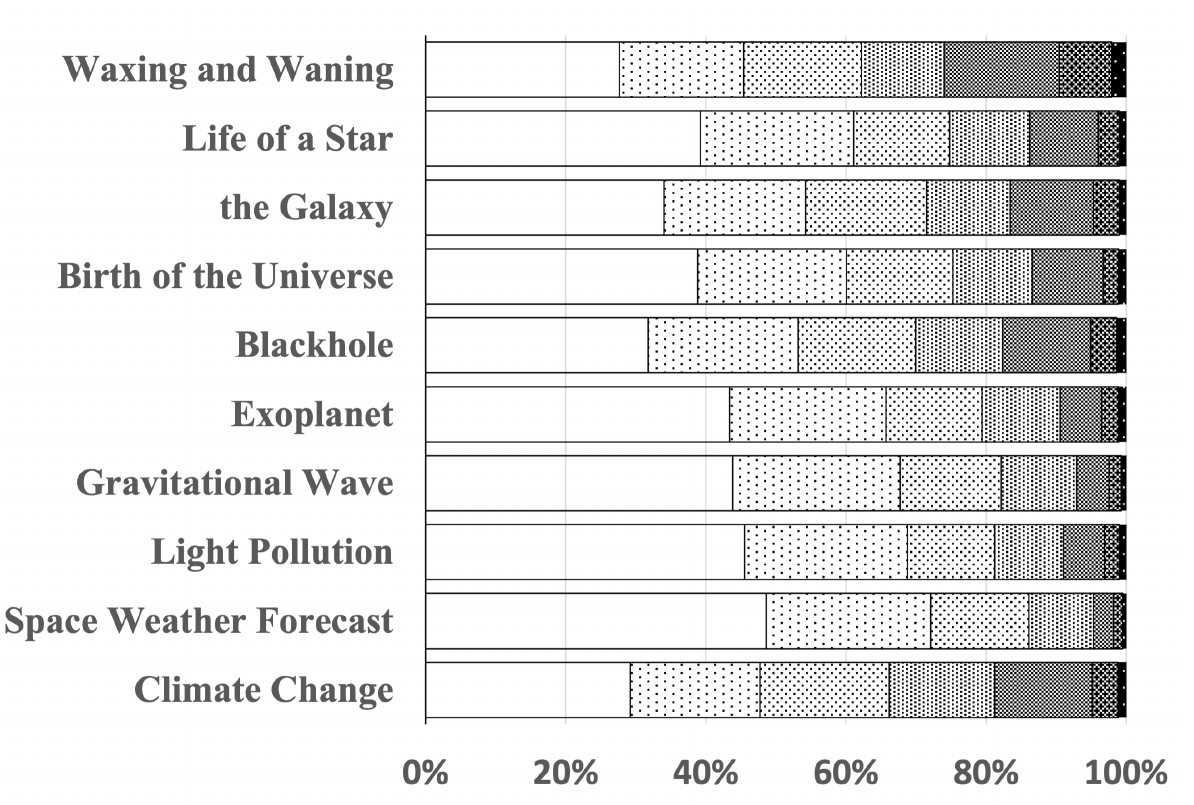}
\raggedright
\caption{Response of Q6, "Can you explain the concepts related to the stars and the Universe?" From left to right: "I can not explain at all," "I can not explain," "I can not explain very much," "I can not say either way," "I can explain," "I can explain a little," "I can explain very much."}
 \label{Q6}
\end{figure}

\begin{figure}[hbtp]
 \centering
 \includegraphics[width=80mm]
      {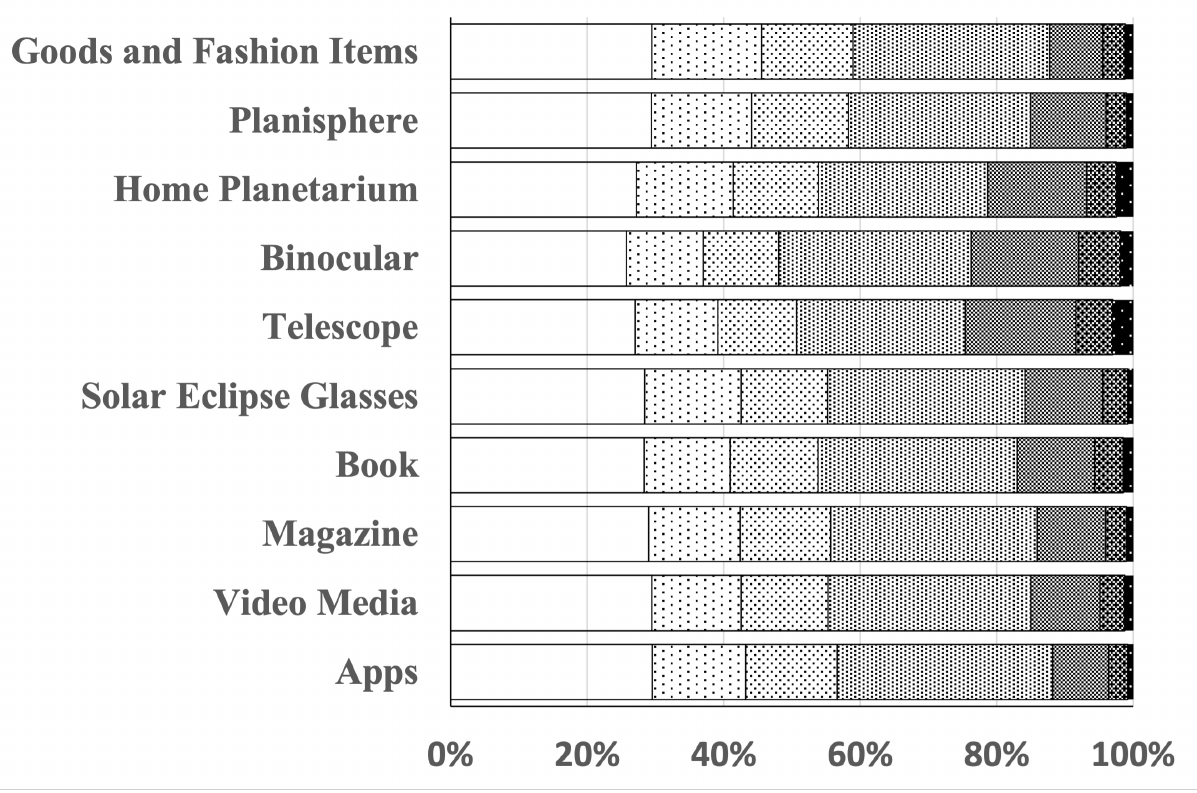}
\raggedright
\caption{Response of Q9, "How much of the items do you want?" From left to right: "I do not want it at all," "I do not want it," "I do not want it very much," "I do not care either way," "I want it," "I want it a little," "I want it very much."}
 \label{Q9}
\end{figure}

\begin{figure}[hbtp]
 \centering
 \includegraphics[width=80mm]
      {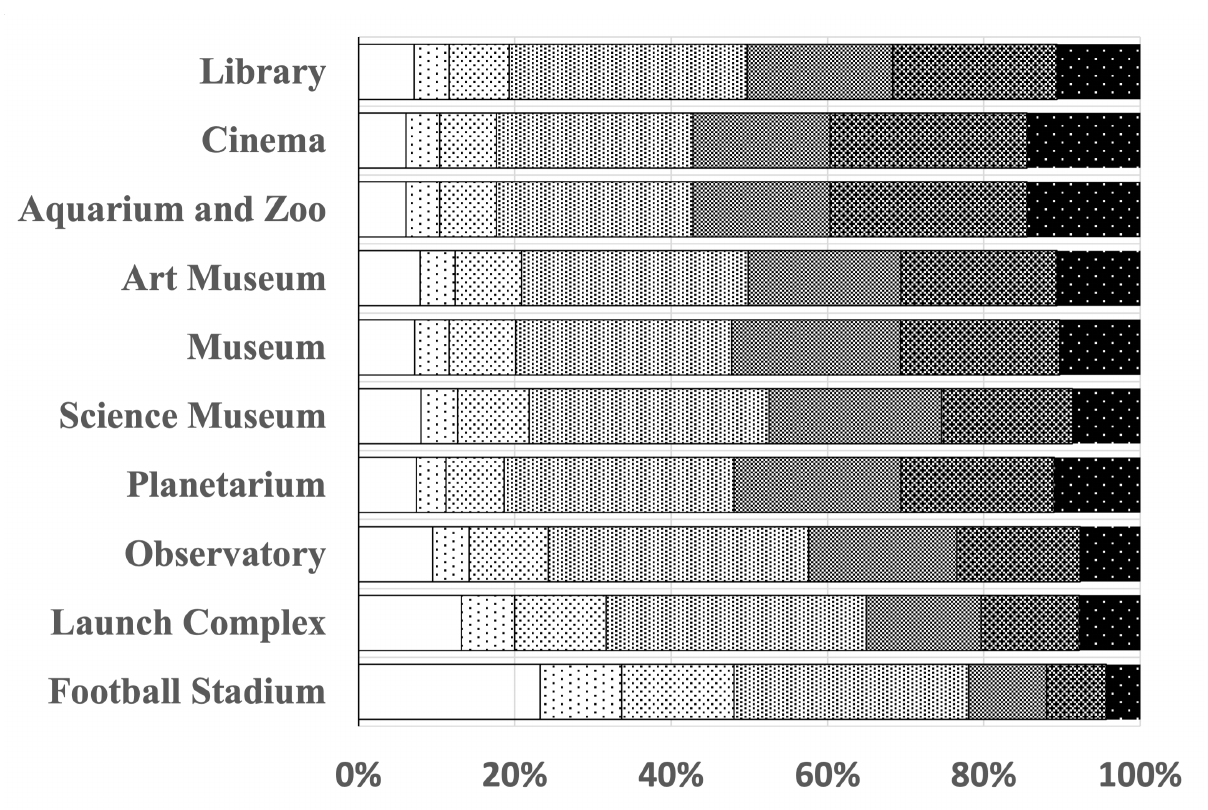}
\raggedright
\caption{Response of Q11, "How much would you like to visit the facilities?" From left to right: "I do not want at all," "I do not want," "I do not want very much," "I do not care either way," "I want ," "I want a little," "I want very much."}
 \label{Q11}
\end{figure}

\begin{figure}[hbtp]
 \centering
 \includegraphics[width=80mm]
      {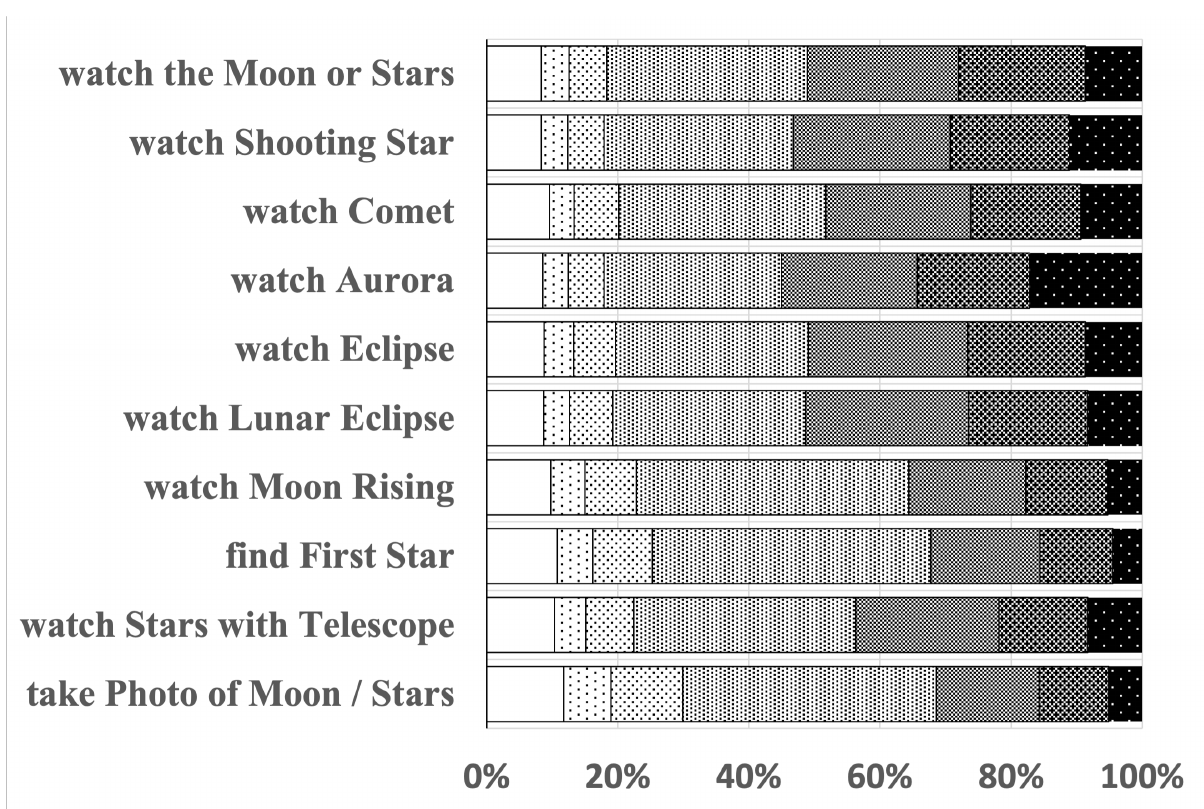}
\raggedright
\caption{Response of Q13, "How much do you want to try the activities?" From left to right: "I do not want at all," "I do not want," "I do not want very much," "I do not care either way," "I want ," "I want a little," "I want very much."}
 \label{Q13}
\end{figure}

\begin{figure}[hbtp]
 \centering
 \includegraphics[width=80mm]
      {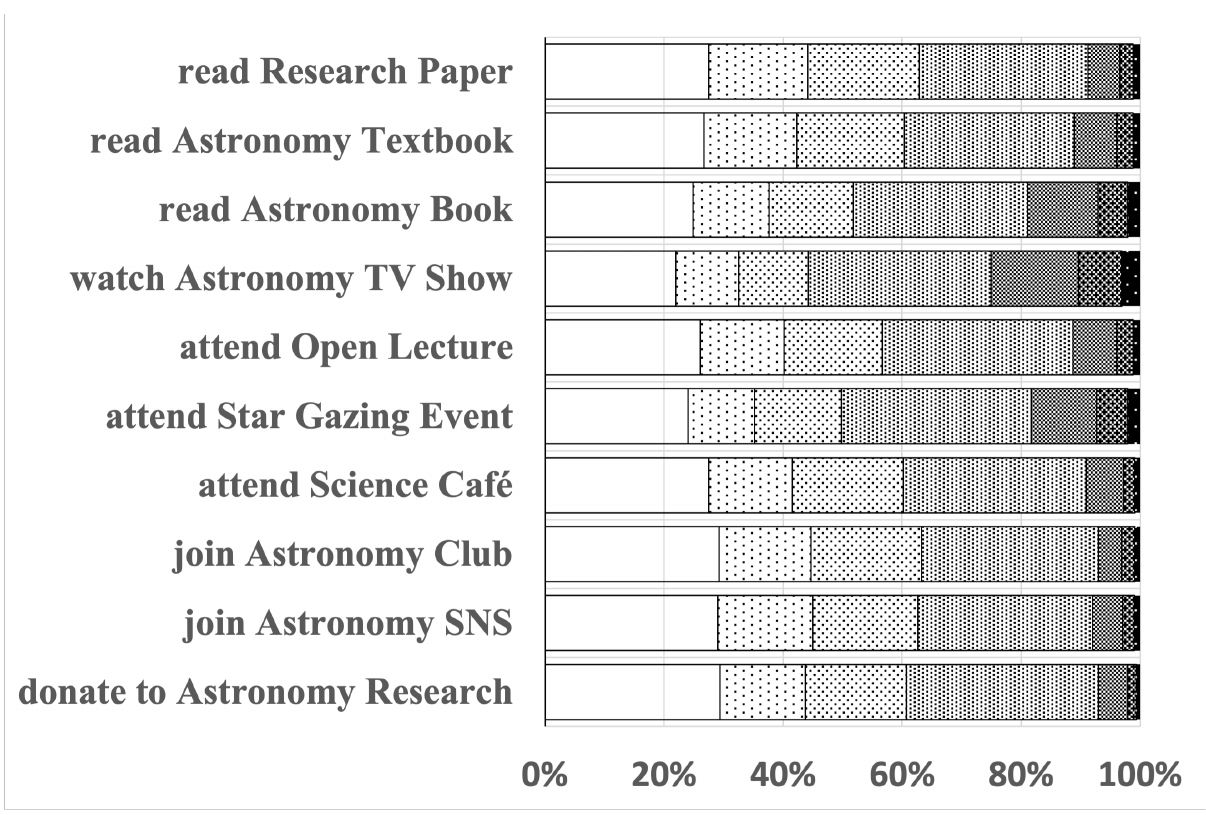}
\raggedright
\caption{Response of Q15, "How much do you want to try the actions?" From left to right: "I do not want at all," "I do not want," "I do not want very much," "I do not care either way," "I want ," "I want a little," "I want very much."}
 \label{Q15}
\end{figure}

%Q1-3
Response to the first three items (Q1--Q3) which are prepared for measuring degree of interest for science and technology is shown in tables \ref{Q1}--\ref{Q3}. Response to the remaining 12 items (Q4--Q15) is shown in tables \ref{Q5}--\ref{Q14} and figures \ref{Q4}--\ref{Q15}.

\begin{table}
\caption{Response to Q1 and Q${_E}$1, "How much are you interested in science and technology?"}%
\label{Q1}
\begin{tabular}{ccc}  
\hline\noalign{\vskip3pt} 
Option & GP\footnotemark[$*$] & EPO\footnotemark[$\dag$] \\  [2pt] 
\hline\noalign{\vskip3pt} 
very interested  & 7.8\% & 66.3\%\\  [2pt] 
quite interested  & 26.1\% & 29.2\%\\  [2pt] 
neither interested nor disinterested  & 29.5\% & 4.5\%\\  [2pt] 
not very interested  & 15.2\% & 0\%\\  [2pt] 
not interested at all & 15.2\% & 0\%\\  [2pt] 
don't know & 6.4\% & 0\%\\  [2pt] 
\hline\noalign{\vskip3pt} 
\end{tabular}\label{table:extramath}
\begin{tabnote}
\footnotemark[$*$] the general public. \\
\footnotemark[$\dag$] the EPO people.
\end{tabnote}
\end{table}

\begin{table}
\caption{Response to Q2 and Q${_E}$2, "Do you actively search for information about science and technology?"}%
\label{Q2}
\begin{tabular}{ccc}  
\hline\noalign{\vskip3pt} 
Option & GP & EPO \\  [2pt] 
\hline\noalign{\vskip3pt} 
Yes  & 21.2\% & 91.0\%\\  [2pt] 
No  & 65.3\% & 5.6\%\\  [2pt] 
don't know & 13.6\% & 3.4\%\\  [2pt] 
\hline\noalign{\vskip3pt} 
\end{tabular}\label{table:extramath}
%\begin{tabnote}
%Q2: {\it Do you actively search for information about science and technology?}
%\end{tabnote}
\end{table}

\begin{table}
\caption{Response to Q3 and Q${_E}$3, "When you have looked for information  science about and technology in the past, have you generally been able to find what you were looking?"}%
\label{Q3}
\begin{tabular}{ccc}  
\hline\noalign{\vskip3pt} 
Option & GP & EPO \\  [2pt] 
\hline\noalign{\vskip3pt} 
\begin{tabular}{c} Yes, and it tends to be \\ easy to understand\end{tabular} & 13.2\% & 59.6\%\\  [2pt] 
\begin{tabular}{c} Yes, but it is often \\difficult to understand\end{tabular} & 30.0\% & 33.7\%\\  [2pt] 
\begin{tabular}{c} No, I often can’ t find \\what I am looking for\end{tabular} & 16.3\% & 2.8\%\\  [2pt] 
don't know & 40.6\% & 3.9\%\\  [2pt] 
\hline\noalign{\vskip3pt} 
\end{tabular}\label{table:extramath}
\end{table}

\begin{table}
\caption{Response to Q5, "Have you studied the Universe using any of the learning opportunities?"}%
\centering
\label{Q5}
\begin{tabular}{cccc}  
\hline\noalign{\vskip3pt} 
Item & Yes & No & don't know\\  [2pt] 
\hline\noalign{\vskip3pt} 
Early Childhood Education & 6.2\% & 64.0\% & 29.9\%\\  [2pt] 
Primary Education & 35.2\% & 43.8\% & 21.1\%\\  [2pt] 
Secondary Education & 38.4\% & 42.1\% & 19.5\%\\  [2pt] 
Higher Education & 22.5\% & 58.0\% & 19.6\%\\  [2pt] 
Postgraduate Education & 3.7\% & 78.7\% & 17.7\%\\  [2pt] 
Learn from Family & 11.4\% & 71.0\% & 17.7\%\\  [2pt] 
Lifelong Learning & 8.1\% & 74.3\% & 17.7\%\\  [2pt] 
\hline\noalign{\vskip3pt} 
\end{tabular}\label{table:extramath}
\end{table}
 
\begin{table}
\caption{Response to Q7, "How do you find information about the Universe?"}%
\centering
\label{Q7}
\begin{tabular}{cccc}  
\hline\noalign{\vskip3pt} 
Item & Yes & No & don't know\\  [2pt] 
\hline\noalign{\vskip3pt} 
Website  & 61.1\% & 25.4\% & 13.5\%\\  [2pt] 
Apps & 8.9\% & 72.2\% & 18.9\%\\  [2pt] 
SNS & 17.7\% & 65.0\% & 17.4\%\\  [2pt] 
Online Video Contents & 28.1\% & 54.6\% & 17.3\%\\  [2pt] 
Book and Magazine & 32.6\% & 50.4\% & 17.1\%\\  [2pt] 
Newspaper & 18.0\% & 63.7\% & 18.3\%\\  [2pt] 
Radio & 8.3\% & 73.6\% & 18.1\%\\  [2pt] 
TV Program & 36.7\% & 47.2\% & 16.1\%\\  [2pt] 
Friend and Family & 22.8\% & 58.3\% & 19.0\%\\  [2pt] 
Expert & 8.6\% & 70.4\% & 21.1\%\\  [2pt] 
\hline\noalign{\vskip3pt} 
\end{tabular}\label{table:extramath}
\end{table}

\begin{table}
\caption{Response to Q8, "Please select the items that you own."}%
\label{Q8}
\begin{tabular}{cc}  
\hline\noalign{\vskip3pt} 
Item & Selected\\  [2pt] 
\hline\noalign{\vskip3pt} 
\begin{tabular}{c} Goods and Fashion Items\\ with Starry Design\end{tabular} & 7.1\%\\  [2pt] 
Planisphere & 7.6\%\\  [2pt] 
Home Planetarium & 2.8\%\\  [2pt] 
Binocular & 15.7\%\\  [2pt] 
Solar Eclipse Glasses & 5.8\%\\  [2pt] 
Telescope & 4.9\%\\  [2pt] 
Book & 6.7\%\\  [2pt] 
Magazine & 3.7\%\\  [2pt] 
Video Media  & 2.4\%\\  [2pt] 
Apps & 2.6\%\\  [2pt] 
I do not have any of them & 70.6\%\\  [2pt] 
\hline\noalign{\vskip3pt} 
\end{tabular}\label{table:extramath}
\end{table}
  
\begin{table}
\caption{Response to Q10, "Please select the facilities you have visited since you became an adult (over 18 years old)."}%
\label{Q10}
\begin{tabular}{cc}  
\hline\noalign{\vskip3pt} 
Facility & Selected\\  [2pt] 
\hline\noalign{\vskip3pt} 
Library & 60.2\%\\  [2pt] 
Cinema & 68.7\%\\  [2pt] 
Aquarium and Zoo & 64.9\%\\  [2pt] 
Art Museum  & 51.2\%\\  [2pt] 
Museum & 45.6\%\\  [2pt] 
Science Center & 34.3\%\\  [2pt] 
Planetarium & 38.0\%\\  [2pt] 
Observatory & 10.3\%\\  [2pt] 
Launch Complex   & 2.2\%\\  [2pt] 
Football Stadium & 18.0\%\\  [2pt] 
I haven't visited any of them & 18.6\%\\  [2pt] 
\hline\noalign{\vskip3pt} 
\end{tabular}\label{table:extramath}
\end{table}

\begin{table}
\caption{Response to Q12, "Which of the activities have you done consciously?"}%
\label{Q12}
\begin{tabular}{cc}  
\hline\noalign{\vskip3pt} 
Activity & Selected\\  [2pt] 
\hline\noalign{\vskip3pt} 
watch the Moon or Stars & 46.4\%\\  [2pt] 
watch Shooting Star & 29.0\%\\  [2pt] 
watch Comet & 11.7\%\\  [2pt] 
watch Aurora  & 4.1\%\\  [2pt] 
watch Lunar Eclipse & 39.2\%\\  [2pt] 
watch Solar Eclipse & 35.2\%\\  [2pt] 
watch Moon Rising & 11.4\%\\  [2pt] 
find Evening Star & 13.1\%\\  [2pt] 
watch Stars with Telescope  & 10.4\%\\  [2pt] 
take Photo of Moon or Stars & 11.8\%\\  [2pt] 
I haven't done any of them & 40.0\%\\  [2pt] 
\hline\noalign{\vskip3pt} 
\end{tabular}\label{table:extramath}
\end{table}

\begin{table}
\caption{Response to Q14 and Q$_{E}$5, "Please select the actions that you have done."}%
\centering
\label{Q14}
\begin{tabular}{ccc}  
\hline\noalign{\vskip3pt} 
Activity & GP & EPO\\  [2pt] 
\hline\noalign{\vskip3pt} 
read Research Paper of Astronomy & 1.9\% & 48.3\%\\  [2pt] 
read Astronomy Textbook & 3.6\% & 66.3\%\\  [2pt] 
read Astronomy Book & 11.1\% & 97.8\%\\  [2pt] 
watch Astronomy TV Show & 16.0\% & 96.1\%\\  [2pt] 
attend Star Gazing Event & 5.0\% & 93.2\%\\  [2pt] 
attend Astronomy Open Lecture & 2.0\% & 88.2\%\\  [2pt] 
attend Science Cafe & 2.0\% & 70.5\%\\  [2pt] 
join Astronomy Club & 1.9\% & 57.9\%\\  [2pt] 
join Astronomy SNS & 2.0\% & 68.5\%\\  [2pt] 
donate to Astronomy Research & 0.7\% & 21.3\%\\  [2pt] 
I haven't done any of them & 76.3\% & N/A\\  [2pt] 
\hline\noalign{\vskip3pt} 
\end{tabular}\label{table:extramath}
\end{table}

\subsubsection{Survey of EPO people}

%socio
The average age of the respondents is 52.7 years old. Note that we asked generations in 10-year age brackets to the EPO people, not actual age, which included some indeterminacy. 4.5\% of the respondents are researchers (astronomy, education, science communication, etc.), 37.1\% are professional communicators (teachers, staff of planetariums, science centers and public observatories, etc.), 46.1\% are citizen communicators (amateur astronomers, volunteers, etc.) and 12.4\% are others (active participant). 

\begin{figure}[hbtp]
 \centering
 \includegraphics[width=80mm]
      {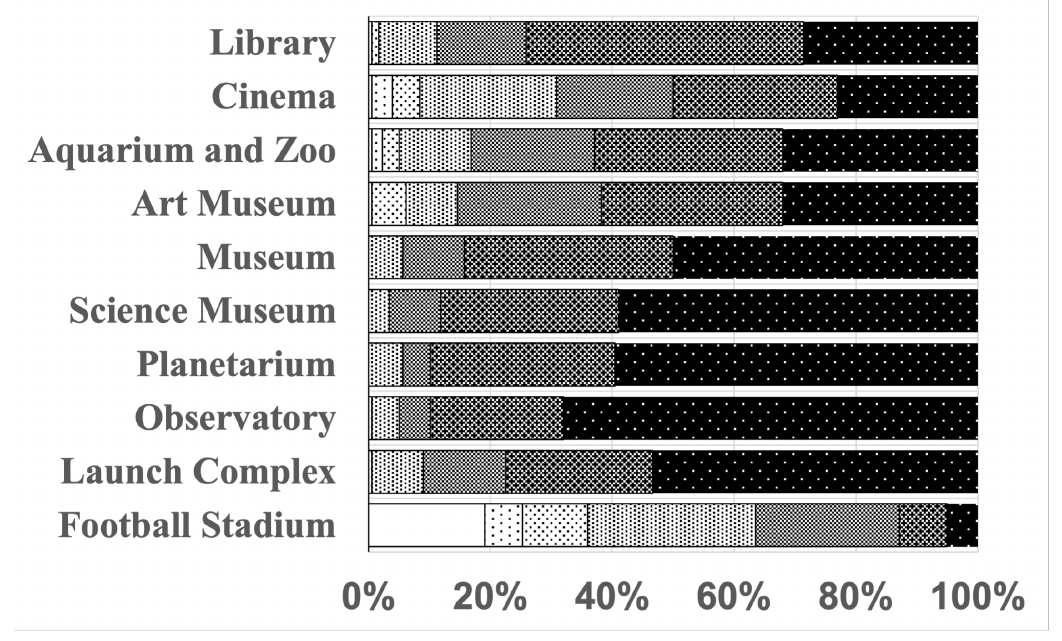}
\raggedright
\caption{Response of Q$_{E}$4, ”How much would you like to visit the facilities?” From left to right: ”I do not want at all,” ”I do not want,” ”I do not want very much,” ”I do not care either way,” ”I want,” ”I want a little,” ”I want very much.”}
 \label{QE4}
\end{figure}

%Q1-3
Response to the first three items (Q$_{E}$1-Q$_{E}$3) which are prepared for measuring degree of interest for science and technology is shown in tables \ref{Q1}-\ref{Q3}. Response to the Q$_{E}$4 is shown in figure \ref{QE4} and response to the Q$_{E}$5 is shown in table \ref{Q14}.

%科学への関心度
\subsection{Interest in Science and Technology}

%VSEGとの
We classified the respondents into six segments (see table \ref{VSEG}, the segment number is ordered from more interest to less interest) using by the VSEG method. According to a report of Framework for Broad Public Engagement in Science, Technology and Innovation Policy (PESTI)\footnote{A research project in Japan aimed at policy formation reflecting the public's needs for science and technology. They reported a distribution of the segments of Japanese public based on a public opinion poll conducted by PESTI in 2013; https://www.nistep.go.jp/research/scisip/data-and-information-infrastructure/pesti-data}, the six segments can be divided into three groups based on their level of interest in science and technology. The combinations are (1) segment 2 and 3 as “group of people with interest in science and technology,” (2) segment 1, 6, and 4 as “people with potential interest in science and technology,” and (3) segment 5 as “people with low interest in science and technology.” In our survey for the general public, we found group (1) is \ch{22.2}\%, group (2) is \ch{43.5}\%, and group (3) is \ch{34.3}\%. For the EPO people, we found group (1) is 93.4\%, group (2) is 6.5\%, and group (3) is 0\%. The numbers add up to 100\% because we exclude respondents who cannot be classified.

\begin{table}
\caption{Segmentation of interest in science and technology.}%
\label{VSEG}
\begin{tabular}{ccc}  
\hline\noalign{\vskip3pt} 
VSEG & GP & EPO \\  [2pt] 
\hline\noalign{\vskip3pt} 
2 & 8.6\% & 56.7\%\\
3 & 10.0\% & 33.1\%\\ 
1 & 11.2\% & 2.8\%\\ 
6 & 2.2\% & 0.6\%\\ 
4 & 23.1\% & 2.8\%\\ 
5 & 28.8\% & 0\%\\ 
NC\footnotemark[$*$] & 15.9\% & 3.9\%\\ 
\hline\noalign{\vskip3pt} 
\end{tabular}\label{table:extramath}
\begin{tabnote}
\footnotemark[$*$] Respondents who can not be classified because they select the item "I don't know.".
\end{tabnote}
\end{table}

\subsection{Trends in Interest in Astronomy}

In order to clarify trends in interest in the Astronomy, the respondents of general public are divided into three groups (group (1) to ($\mathrm{iii})$) as we showed in \S4.2, and the respondents of EPO people are divided into four groups; (4) researchers, (5) professional communicators, ($\mathrm{vi}$) citizen communicators, and (7) others. Table \ref{GROUP} shows how many are included in each group.

\begin{table}
\caption{Number of the groups.}%
\label{GROUP}
\begin{tabular}{cclc}  
\hline\noalign{\vskip3pt} 
& Gr. & Description & Number \\  [2pt] 
\hline\noalign{\vskip3pt} 
GP & 1 & with interest & 372\\
& 2 & with potential interest & 733\\ 
& 3 & with low interest & 576\\ \hline 
EPO & 4 & researchers & 8\\ 
& 5 & professional communicators & 65 \\ 
& 6 & citizen communicators & 78 \\ 
& 7 & others & 27\\ 
\hline\noalign{\vskip3pt} 
\end{tabular}\label{table:extramath}
\end{table}

\begin{figure}[hbtp]
 \centering
 \includegraphics[width=80mm]
      {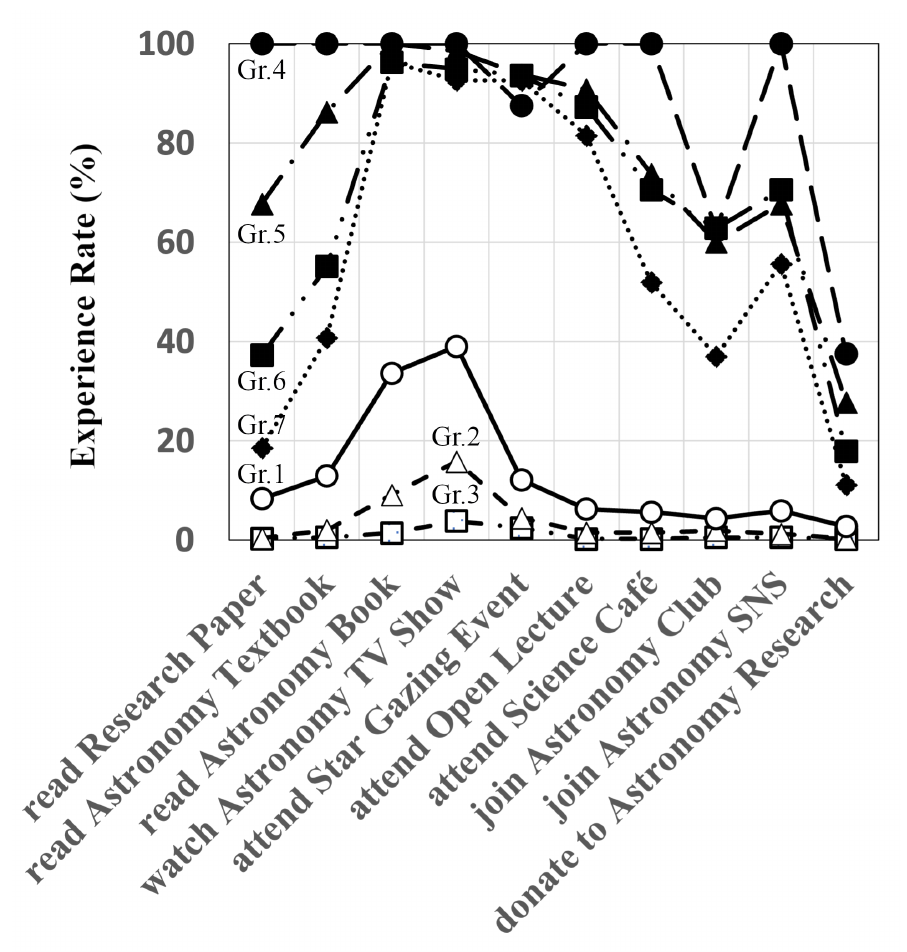}
\raggedright
\caption{Experience rate of the activities. Vertical axis unit is percentage. The general public divided into three groups; people with interest in science and technology (open circle with solid line, group (1)), people with potential interest in science and technology (open triangle with dashed line, group (2)), and people with low interest in science and technology (open square with dot--dashed line, group (3)). The EPO people divided into four groups; researchers (astronomy, education, science communication: filled circle with long dashed line, group (4)), professional communicators (teachers, staff of planetariums, science centers and public observatories: filled triangle with dot--long dashed line, group(5)), citizen communicators (amateur astronomers, volunteers: filled square with dot--dot-long dashed line, group (6)) and others (participant of activities: filled diamond with dotted line). The "astronomy book" here is intended to be an enlightening book for the general public.}
 \label{Q15_COMPARE}
\end{figure}

\begin{figure}[hbtp]
 \centering
 \includegraphics[width=80mm]
      {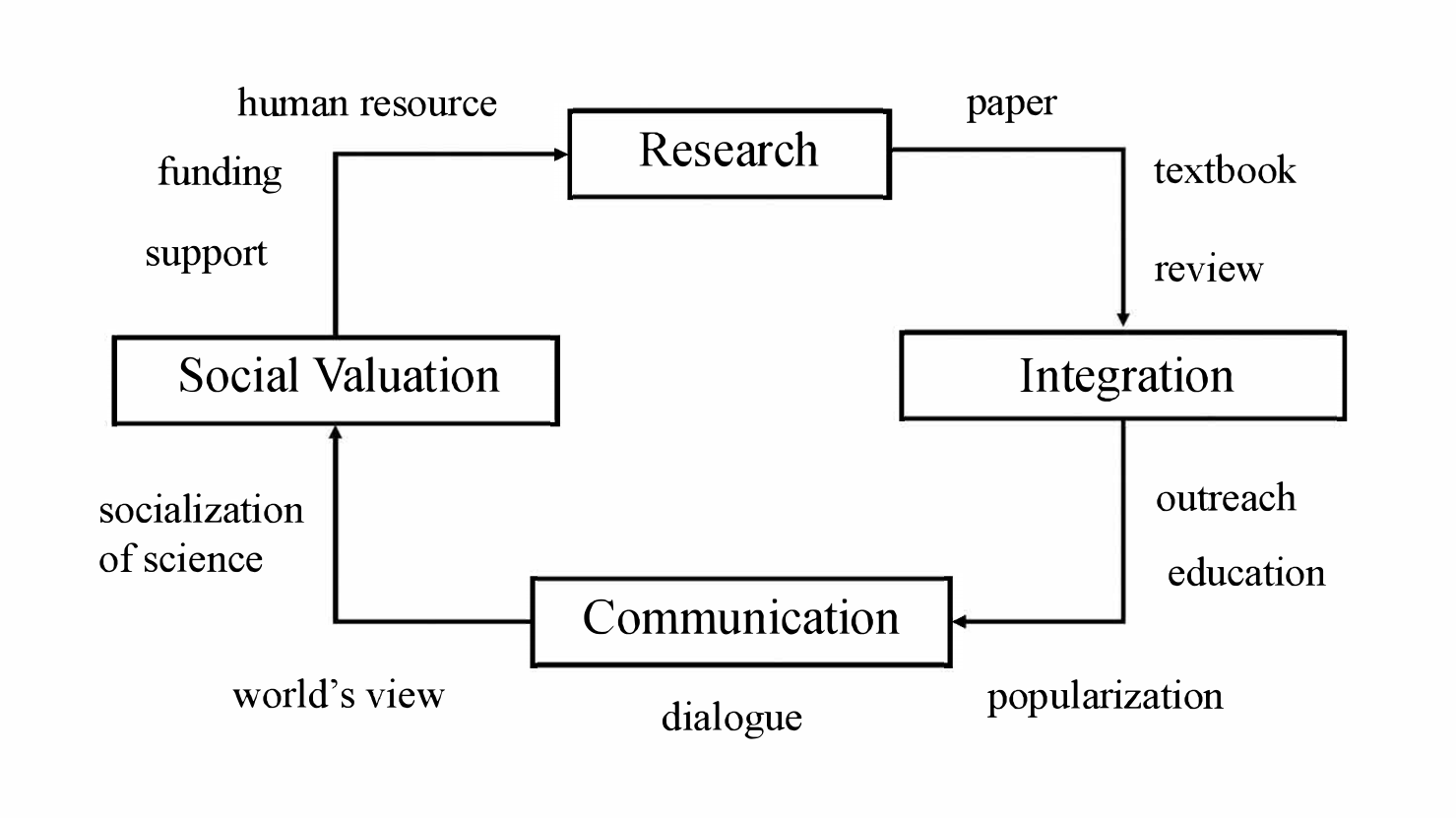}
\raggedright
\caption{Conceptual framework of the circulation model of knowledge proposed by \cite{taka2014}. The model consists of four sections: (a) Research; researchers produce scientific papers in each sub-field of astronomy, (b) Integration; the papers are reviewed and digested from higher point of views and packaged into textbooks, general books, TV programs, and so on, (c) Communication; the packaged knowledge is absorbed by people in various ways and interwoven with people's views of the world, and (d) Social Valuation; as a result, people find their own views of astronomy and the aggregate of those values become the social valuation of astronomy. This valuation encourages the next step of research and the circulation will develop astronomy and fertilize the society.}
 \label{model}
\end{figure}

Figure \ref{Q15_COMPARE} shows how the respondents interacted with astronomy. The activities mentioned in the figure are linked to the circulation model of knowledge of Astronomy (see figure \ref{model}, cf. \cite{taka2014, taka2018}), which describes how the fruits of the Astronomy are transmitted to society and then back to promote Astronomy. The experience rates of "read Astronomy Book," "watch Astronomy TV Show," "attend Star Gazing Event" are over 90\% for the EPO people and clearly different from those of the general public. In contrast, there is variation within the EPO for "read Research Paper," "read Astronomy Textbook," "donate to Astronomy Research," which appears to distribute continuously, including among the general public. If we look at the general population, we find that the higher the level of interest in science in "read Astronomy Book" and "watch Astronomy TV Show," the more likely they are to be experienced.

\begin{figure}[hbtp]
 \centering
 \includegraphics[width=80mm]
      {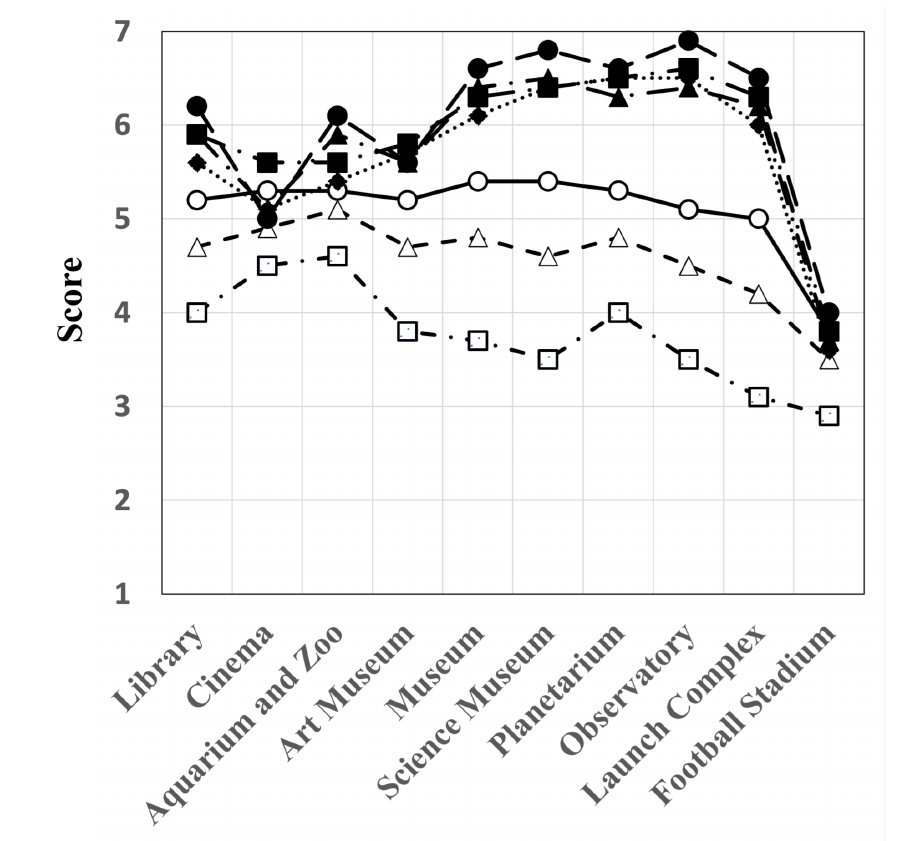}
\raggedright
\caption{Score that shows how much the respondents want to visit the facilities. The vertical axis represents the average of the scores on a 7-point scale from passive to aggressive. Symbols and lines are the same as figure \ref{Q15_COMPARE}}
 \label{Q11_COMPARE}
\end{figure}

Figure \ref{Q11_COMPARE} shows how much the respondents want to visit facilities. The EPO people are obviously more eager to visit Library, Museum, Science Center, Planetarium, Observatory, and Launch Complex than the general public. On the other hand, there are less differences in the eagerness in visiting Cinema and Football Stadium between the EPO people and the general public, except for group (3).

%%一般

\section{Discussion}

\subsection{Public Attitude toward Astronomy}

We discuss what kind of interest the general public in Japan has in astronomy based on the data obtained. 

\begin{table}
\caption{Comparison of group ratios with previous studies.}%
\label{COMPARE_VSEG}
\begin{tabular}{ccccc}  
\hline\noalign{\vskip3pt} 
& Gr.(1) & Gr.(2) & Gr.(3) & N\\  [2pt] 
\hline\noalign{\vskip3pt} 
this study & \ch{22.2}\% & \ch{43.5}\% & \ch{34.3}\% & 1681\\
Goto\footnotemark[1] & 30.5\% & 51.4\% & 18.1\% & 755\\
PESTI\footnotemark[2] & 16.1\% & 61.4\% & 22.6\% & 846\\
Ikkatai\footnotemark[3] & 26.7\% & 48.5\% & 24.9\% & 914\\
\hline\noalign{\vskip3pt} 
\end{tabular}\label{table:extramath}
\begin{tabnote}
\footnotemark[1] \citet{goto2015}\\
\footnotemark[2] Public opinion poll conducted in Japan in 2013, see \S5.2.\\
\footnotemark[3] \citet{ikkatai2022}
\end{tabnote}
\end{table}

First, let us review the statistical validity of our survey results. Since there are about 90 million Japanese aged 20-69, the smallest sample size that would have a 5\% margin of error and 95\% confidence level assuming a 50\% response rate is 384. Notice since the survey was conducted using registered monitors, the true response rate is unknown. We believe that the actual response rate is higher than 50\%, and 50\% is a figure we set as a lower limit. From this point forward, the discussions about the general public will be conducted for each of the three groups categorized based on their level of interest in science and technology (see Table \ref{GROUP}), and the smallest sample size will be in the group (\ch{1}), which includes 372 respondents. Although this number is slightly below the minimum, we recognize that the following discussion is statistically significant.

The typical Japanese attitude toward astronomy presented in the survey data is as follows. They are interested in a variety of things, and nearly 30\% of them has some interest in astronomy (figure \ref{Q4}). They are more interested in climate change and Japanese culture than in astronomy, but are equally interested in art and space exploration, and less interested in mythology, astrology, philosophy, religion, and physics. However, not many people, even those with some interest in astronomy, can explain the terms about astronomy. The rates varies from words to words; a few percent for ""Space Weather Forecast" to 30 percent for "Waxing and Waning" (Figure \ref{Q6}). The majority of the general public think that they have never studied astronomy in any education program (table \ref{Q5}), although some of the astronomy-related topics, such as the apparent motions of the Sun and the mechanism of the phases of the moon, is included in the official curricula of Japanese elementary and junior high school. They get information related astronomy primarily from websites (table \ref{Q7}). They also refer traditional media such as books and magazines, but online media seems to be influential. Binoculars are by far the most common astronomical items owned by these people (table \ref{Q8}). On the other hand, while the ownership rate is low, the desire to own a home planetarium or an astronomical telescope is high (figure \ref{Q9}). They have visited a variety of facilities (table \ref{Q10}). One-third of respondents have visited science centers and planetariums, while only one-tenth of respondents have visited observatories. However, the percentage of respondents who would like to visit these facilities generally accounts for half of the total respondents, which is about the same level of popularity as other commonly visited facilities such as libraries, cinemas, aquariums, zoos, and so on (figure \ref{Q11}). Many of them have had the experience of watching the moon and stars, and astronomical phenomena such as solar and lunar eclipses (table \ref{Q12}). Far fewer have had the experience of watching comets and aurora, watching stars with a telescope, but the desire to do so is very high (figure \ref{Q13}). In terms of the cyclic knowledge model, 76.3\% of them is not involved at all, and reading books and watching TV shows are the most familiar experiences (table \ref{Q14}). Few have ever attended a star gazing party, but many would like to (figure \ref{Q15}). 

The typical Japanese image described here is just our interpretation, and there may well be room for other interpretations. In this sense, it is necessary to verify whether our interpretations are appropriate. However, these verifications are out of the scope of this paper. These verifications will be conducted in future studies. In this study, we only examine whether the data on which this interpretation is based is not anomalous.

One feature of this study is that we use the VSEG method to classify respondents based on their level of interest in science and technology. Table \ref{COMPARE_VSEG} compares the results of this classification with those of previous studies. All of these previous studies were conducted with the general Japanese public. The analyses by \citet{goto2015} and \citet{ikkatai2022} are based on the online surveys, as were this study, but PESTI differed in that it conducted its survey face-to-face. While there are some differences in the age range of the respondents, we do not recognize that this has a significant impact on the results. While there is some variation in the numbers of table \ref{COMPARE_VSEG}, the trend seems to be the same, with the group (2), "potentially interested in science and technology," being the largest group. We are concerned about the slightly larger number (\ch{34.3}\%) of group (3) in our study, but we do not know whether this is due to a systematic reason or not. We proceed discussion under the assumption that at least our respondents are not an obviously unusual group.

\subsection{Difference between the general public and the EPO people}

First, we discuss the statistical significance of the survey for the EPO people. For groups (\ch{5}) and (\ch{6}) in Table\ref{GROUP}, the population is estimated to be in the hundreds\footnote{Estimated based on the number of members of JESPA, JAPOS and JPA to which Japanese education and extension professionals belong.} and thousands\footnote{Estimated from the number of star guide training course graduates, etc.}, respectively, and therefore, a 10\% margin of error and a 90\% confidence level are considered acceptable. For group(7), although the size of the population is unknown, it is estimated to be over one million people\footnote{Estimated from the number of visitors to the planetarium, etc.}, so it is considered to have a 20\% margin of error and a 95\% confidence level. For group (4), the population is estimated to be only a few hundred (cf. \cite{sawa2000}), a 20\% margin of error and a 80\% confidence level are considered acceptable. Furthermore, we did not ask for cooperation on the mailing list of the Astronomical Society of Japan\footnote{https://www.asj.or.jp/en/}, to which the largest number of astronomers in Japan belong. This is because the purpose of this survey was to first investigate the trends among those involved in education and public outreach. Therefore, the "researchers" referred to here are likely to have a strong bias toward those who are strongly interested in education and public outreach. In short, the results of the survey of the EPO people are less statistically significant and biased than those of the survey of the general public, and the following discussion is based on that assumption. 

The difference in attitude toward astronomy between the general public and EPO people is obvious. Comparing figure \ref{Q11} and \ref{QE4}, the percentage of respondents who would like to visit science centers, planetariums, observatories, and launch sites, which are related to astronomy and space, exceeded 90\%, which is significantly different from the general public, where the percentage is generally below 50\%. Table \ref{Q14}, which asks respondents about their experience participating in astronomy-related activity, also shows the general public and the EPO people are quite different.

It is clear that there are differences between them, but is it possible to find continuity between them? Focusing on the level of interest in science and technology, since most of the EPO people are interested in science and technology (93.4\%, see \S4.2), it can be inferred that there is a close relationship between the general public with interest (group (1)) and the EPO peoples. In fact, when comparing the degree to which people want to go to facilities which less related to astronomy such as libraries, cinemas, aquariums, zoos, art museums, and football stadiums, group (1) is consistent with the EPO people (however, museums are also less related to astronomy, but it has a difference in degree). 

Figure \ref{Q15_COMPARE}, which divides the EPO people into four groups based on their attributes, suggests more continuity between group (1) and the EPO people. It is an interesting fact that reading research papers and astronomy textbooks is in order from group (7), which is considered less involved in astronomy, to group (4), which is considered more involved. It is also noteworthy that group (7) comes between group (1) and the rest of the EPO people when focusing on involvement in more interactive events such as attending science cafes and participating in astronomy clubs.

\begin{figure}[hbtp]
 \centering
 \includegraphics[width=80mm]
      {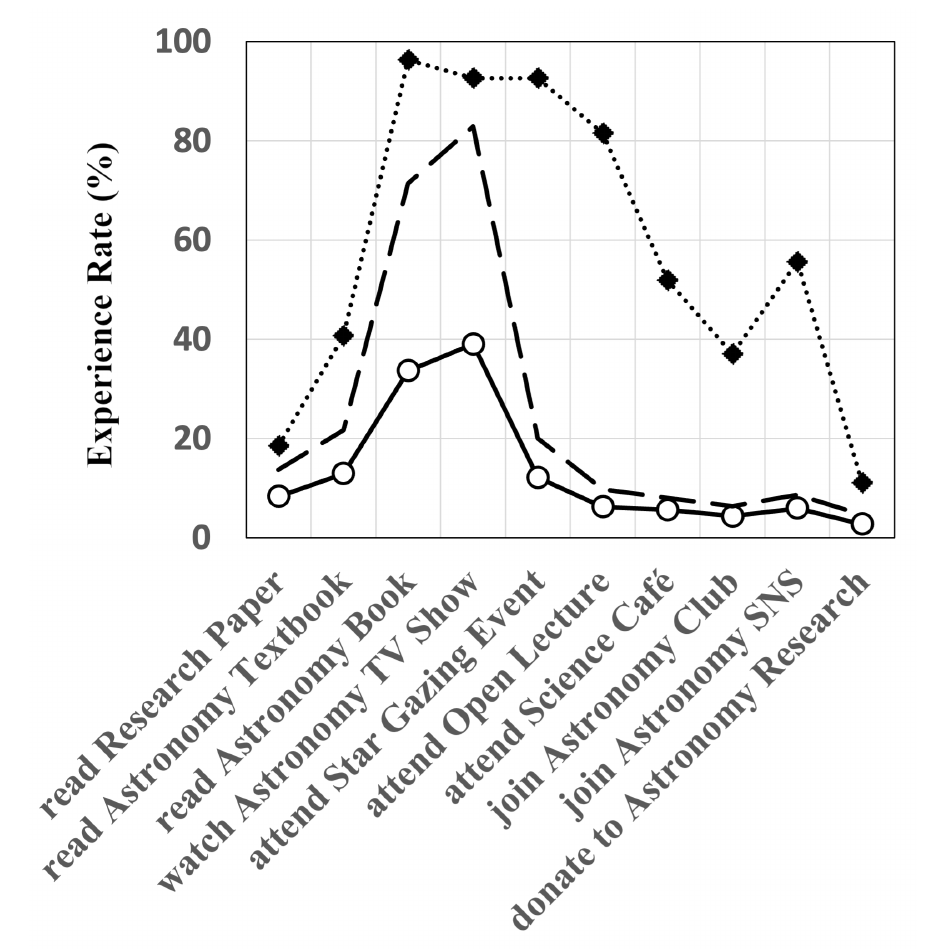}
\raggedright
\caption{Experiment rate of the activities, same as figure \ref{Q15_COMPARE}. We plot only three groups here : group (1) (open circle with solid line), group (7) (active participant: filled diamond with dotted line) and the selected group (1) respondents who have an experience of reading astronomy books or watching astronomy TV shows (long dashed line).}
 \label{Q15_CONT}
\end{figure}

\begin{figure}[hbtp]
 \centering
 \includegraphics[width=80mm]
      {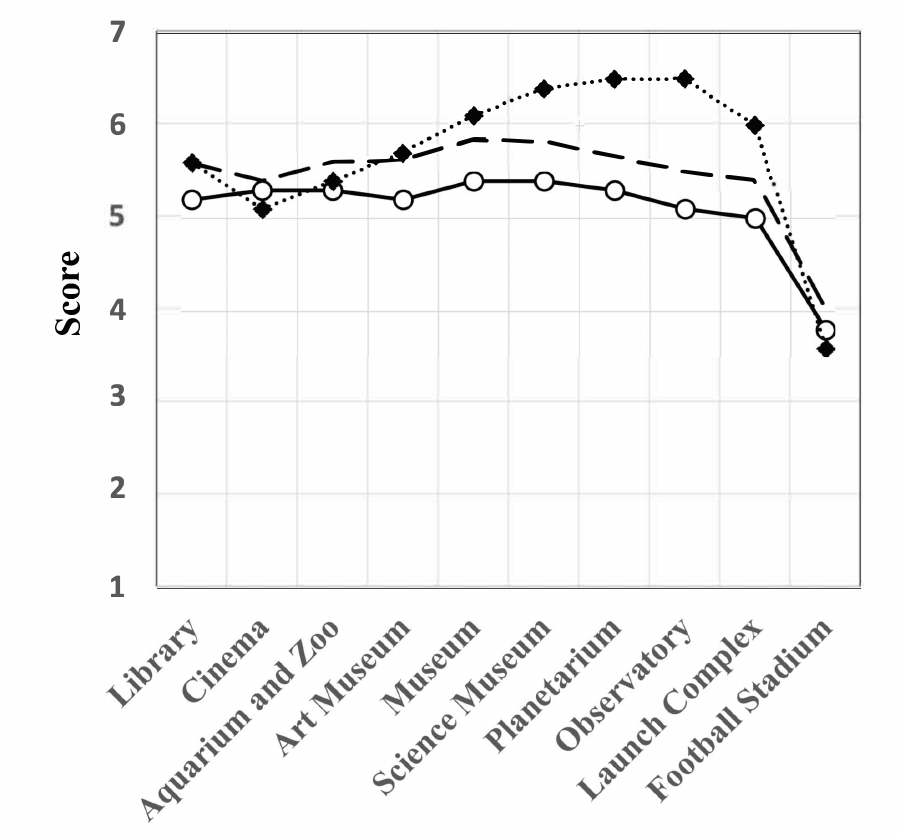}
\raggedright
\caption{Score that shows how much the respondents want to visit the facilities, same as figure \ref{Q11_COMPARE}. Symbols and lines are the same as figure \ref{Q15_CONT}}
 \label{QE4_CONT}
\end{figure}

Figure \ref{Q15_COMPARE} shows that reading astronomy books and watching astronomy TV shows are activities that a relatively large number of people in group (1) have done. Figure \ref{Q15_CONT} plots only those in group (1) who have engaged in one of these two activities ($n=175$). This certainly shows how the experience of one of these two activities precedes the other activities. In a similarly plotted figure \ref{QE4_CONT}, it is consistent to assume that those with either experience are more willing to visit astronomy-related facilities than those in group (1) and closer to group (7). These results suggest a continuum between the general public and the EPO people. These data shows that 
reading books and watching astronomy TV shows have become the easiest entry point into astronomy for the general public. On the other hand, the data also suggests that to become an active participant, one needs to be more aggressive, such as attending a star gazing event.

We would like to confirm the continuity between them in terms of sociodemographic parameters such as age, gender and so on, but it is clear that the survey data of the EPO people is biased so these points cannot be verified. It will be the subject of the next study. In addition, although we discussed this issue only from a quantitative perspective, it would be important to conduct interviews with the EPO people to gain a better understanding of what steps they took to become deeply involved in astronomy. Through such data-based research, we will be able to clarify the relationship between society and astronomy.

\ch{\subsection{How should the data be used?}}

\ch{In what sense would the data we presented be useful? We point out the following three perspectives.}

\ch{First of all, we point out that as a community, the data is an important basis for building a better relationship between astronomy and society. More specifically, it will help in determining what we should do for the above purpose from a strategic perspective. Based on the data shown here, it would be more constructive to discuss the following, but not limited to, issues; desirable relationship between astronomy and society, problem finding, agenda shaping, solution, effective resource allocation, effect measurement.}

\ch{Next, we point out that the data helps us to establish a methodology of communicating astronomy to the public. Through the analysis of the data, we will be able to discover new findings and research questions related to such things. With the exception of limited areas such as school education, research of the relationship between astronomy and the public has lagged far behind the development of astronomy as a science. It is clear that its importance is growing, and it is hoped that studies following this survey will emerge.}

\ch{Finally, we point out that if you are a researcher, the data is useful for you to realize how unique you are. In other words, it tells you how out of touch you are with the general sense of the public. That helps us to communicate with people about the astronomy.}

\ch{The data presented here was obtained by the first survey focusing on the public attitude toward astronomy in Japan, and could be used in a variety of future studies. It should be noted, however, that this data only represent some aspects of the Japanese public's attitudes as of 2023. The attitudes may fluctuate with changes in society. In order to foster a better relationship between astronomy and society, it is important to regularly monitor trends and develop strategies that respond to the trends.}

\section{Summary}

We conducted an online survey of the Japanese general public to investigate their attitudes toward astronomy. This survey of 2,000 respondents of the general public was the first astronomy-specific survey of its kind conducted in Japan. As a result, we were able to gain a quantitative understanding of the general public's interests, experiences, and so on.

We also investigated the differences between the attitudes of the general public and those involved in astronomy education and public outreach. The results showed that while the two groups clearly differ in terms of their involvement in astronomy and the facilities they have visited, there is some continuity between the two groups when categorized by interest in science and technology.

However, the conclusions here need to be validated by other studies. In addition to quantitative surveys such as we conducted in this study it would be helpful to conduct qualitative surveys as well to gain a deeper understanding of the actual situation. Quantitative surveys could also be conducted several more times to estimate the magnitude of systematic uncertainty due to the survey methodology. It is quite possible that the survey results may have changed over time, and multiple independent surveys would be necessary to ascertain such changes. It would also be necessary to use cross tabulation to deeper understand the data obtained. 

\ch{The data presented in this paper could be useful in several ways to promote communicating astronomy to the public.} All data is made publicly available, and we hope that it will be used in various surveys and studies.

\section{Acknowledgement}
This work was supported by JSPS KAKENHI Grant Number 22K03016. We also thank to the TENPLA Project for the funding.

\appendix
\section{Questionnaire Items}
Table 13 shows the questionnaire items for the surveys. Q1-Q15 are items for the survey of general public and Q$_{E}$1-Q$_{E}$5 are items for the survey of the EPO people. The original questionnaire is written in Japanese.

\begin{table*}
\tbl{Questionnaire items used in this study.}{%
\label{Qs}
\begin{tabular}{ll}  
\hline\noalign{\vskip3pt} 
ID & Questionnaire \\  [2pt] 
\hline\noalign{\vskip3pt} 
Q1, Q$_{E}$1  &  How much are you interested in science and technology? \\  [2pt] 
Q2, Q$_{E}$2  &  Do you actively search for information about science and technology? \\  
Q3, Q$_{E}$3  &  When you have looked for information  science about and technology in the past,  \\
& have you generally been able to find what you were looking for? \\  
Q4  &  How interested are you in the items?  \\
opt. & \textit{Mythology, Astrology, Japanese Culture, Philosophy, Religion} \\ 
& \textit{Art, Climate Change, Space exploration, Astronomy, Physics}\\  
Q5  &  Have you studied the Universe using any of the learning opportunities?  \\
opt. & \textit{Early Childhood Education, Primary Education, Secondary Education, Higher Education,}\\  
& \textit{Postgraduate Education, Learn from Family, Lifelong Learning}\\ 
Q6  &  Can you explain the concepts related to the stars and the Universe?  \\
opt. & \textit{Waxing and Waning, Life of a Star, the Galaxy, Birth of the Universe, Blackhole, Exoplanet,}\\ 
& \textit{Gravitational Wave, Light Pollution, Space Weather Forecast, Climate Change}\\ 
Q7  &  How do you find information about the Universe?  \\
opt. & \textit{Website, Apps, SNS, Online Video Contents, Book and Magazine, Newspaper, Radio, TV Program, Friend and Family, Expert}\\ 
Q8  & Please select the items that you own.  \\
opt. & \textit{Goods and Fashion Items with Starry Design, Planisphere, Home Planetarium, Binocular, Telescope,}\\ 
& \textit{Solar Eclipse Glasses, Book, Magazine, Video Media, Apps, I do not have any of them}\\ 
Q9  & How much of the items do you want?  \\
opt. & \textit{same options as Q8}\\ 
Q10, Q$_{E}$4  & Please select the facilities you have visited since you became an adult (over 18 years old). \\
opt. & \textit{Library, Cinema, Aquarium and Zoo, Art Museum, Museum, Science Center, Planetarium, Observatory, }\\ 
& \textit{Launch Complex, Football Stadium, I do not have any of them}\\ 
Q11  &  How much would you like to visit the facilities?  \\
opt. & \textit{same options as Q10}\\ 
Q12  & Which of the activities have you done consciously? \\
opt. & \textit{watch the Moon or Stars, watch Shooting Star, watch Comet, watch Aurora, watch Lunar Eclipse, watch Solar Eclipse, }\\ 
& \textit{watch Moon Rising, find Evening Star, watch Stars with Telescope, take Photo of Moon or Stars,}\\ 
& \textit{I haven't done any of them}\\ 
Q13  & How much do you want to try the activities?  \\
opt. & \textit{same options as Q12}\\ 
Q14, Q$_{E}$5  & Please select the actions that you have done. \\
opt. & \textit{read Research Paper of Astronomy, read Astronomy Textbook, read Astronomy Book, watch Astronomy TV Show,}\\ 
& \textit{attend Star Gazing Event, attend Astronomy Open Lecture, attend Science Cafe, join Astronomy Club,}\\ 
& \textit{join Astronomy SNS, donate to Astronomy Research, I haven't done any of them}\\ 
Q15  & How much do you want to try the actions?  \\
opt. & \textit{same options as Q14}\\ 
\hline\noalign{\vskip3pt} 
\end{tabular}}
\end{table*}

\end{document}